\documentclass{emulateapj}

\newcommand{\eqref}[1] {equation $($\ref{#1}$)$}
\def\beq{\begin{equation}}
\def\eeq{\end{equation}}
\def\beqn{\begin{eqnarray}}
\def\eeqn{\end{eqnarray}}

\def\cl{C_{\ell}}

\def\OL{\Omega_\Lambda}
\def\Om{\ensuremath{\Omega_{\mathrm{m}}}}

\def\msol{\ensuremath{M_\odot}}
\def\l{\left}
\def\r{\right}

\def\2gcm{\textrm{g cm$^{-2}$}}

\def\rcr{\rho_{\mathrm{crit}}}

\def\modu#1{\l |{#1}\r |}
\def\av#1{\l \langle{#1}\r \rangle}
\def\hmpc{\:{h}^{-1}\mathrm{Mpc}}

\def\Sig{\Sigma}

\def\k{\kappa}

\def\hunit{\ensuremath{\mathrm{km}{\mathrm{s}^{-1}} \mathrm{Mpc}^{-1}}}
\def\H0{\ensuremath{\mathrm{H}_0}}
\def\nhat{\hat{\bmm{n}}}

\def\nn{\nonumber}

\def\fsky{f_{\mathrm{sky}}}
\newcommand{\E}[1]{\times 10^{#1}}
\newcommand{\bmm}[1]{{\mathbf{#1}}}

\newcommand{\bm}[1]{\ensuremath{\mbox{\boldmath $#1$}}}

\def\R200{\ensuremath{r_{200}}}
\def\M200{\ensuremath{M_{200}}}

\def\healpix{HEALPix }

\def\pixarea{\ensuremath{\Omega_\mathrm{pix}\: }}
\def\ncap{\ensuremath{N_\mathrm{cap}\: }}

\def\caparea{\Omega_\mathrm{cap}}
\def\tildecaparea{\tilde{\Omega}_\mathrm{cap}}
\def\capangle{\theta_\mathrm{cap}}
\def\steradian{\mathrm{sr}}
\def\mslice{M_\mathrm{slice}}

\def\nsidenow{$4096\:$}
\def\maxz{4.0}
\def\angresnow{\ensuremath{0.\arcmin 896 \:}}
\def\lm{{\ell m}}
\def\lmprime{{\ell' m'}}
\def\lone{{\ell_1}}
\def\ltwo{{\ell_2}}

\def\lthree{{\ell_3}}
\def\lmone{{\lone m_1}}

\def\lmax{{\ell_{\mathrm{max}}}}
\def\etheta{\mathbf{e}_{\theta}}
\def\ephi{\mathbf{e}_{\phi}}

\newcommand\set[1]{\{#1\}}

\graphicspath{{figures/}}

\begin{document}
\title{A Large Sky Simulation of the Gravitational Lensing of the
Cosmic Microwave Background}
\author{Sudeep Das \& Paul Bode}
\affil{Princeton University Observatory\\
Peyton Hall, Ivy Lane, Princeton, NJ 08544 USA}
\email{sudeep@astro.princeton.edu,bode@astro.princeton.edu}
\shortauthors{Das \& Bode}
\shorttitle{Large Sky CMB Lensing Simulation}
\begin{abstract}
Large scale structure deflects cosmic microwave background (CMB) photons. Since large angular scales in the large scale structure contribute significantly to the gravitational lensing effect, a realistic simulation of CMB lensing requires a sufficiently large sky area. We describe simulations that include these effects, and present both effective and multiple plane ray-tracing versions of the algorithm, which employs spherical harmonic space and does not use the flat sky approximation. We simulate lensed CMB maps with an angular resolution of $\sim 0.\arcmin9$. The angular power spectrum of the simulated sky agrees well with analytical predictions.  Maps generated in this manner are a useful tool for the analysis and interpretation of upcoming  CMB experiments such as PLANCK and ACT.
\end{abstract}

\section{Introduction }
While the current generation of CMB experiments have  had
a significant impact on cosmology by helping to establish a
standard paradigm for cosmology \citep{2003ApJS..148..175S,
2007ApJS..170..377S}, the upcoming generation of CMB experiments still
has the potential to provide novel new insights into cosmology.
PLANCK\footnote{\url{http://www.rssd.esa.int/index.php?project=Planck}} and ground based experiments, such as the Atacama Cosmology
Telescope (ACT)\footnote{\url{http://www.physics.princeton.edu/act}}, will be mapping the CMB sky with significantly
higher angular resolution than ever before.
Secondary anisotropies on small
angular scales encode important information about the
late time interaction of CMB photons with structure in the Universe.
One of the most basic of these interactions is the gravitational
effect of the large scale structure potentials
deflecting the paths of the photons, an effect justifiably
referred to as the Gravitational Lensing of the CMB.

The effect of gravitational lensing can be thought of as
a remapping of the unlensed CMB field by a line-of-sight
averaged deflection field \citep[for a recent review,
see][]{2006PhR...429....1L}. Therefore, lensing does not change
the one-point properties of the CMB. However, it does modify
the two and higher-point statistics, and generates non-Gaussianity
\citep{1996ApJ...463....1S,1999PhRvD..59l3507Z,2000PhRvD..62f3510Z}.
Although the typical deflection suffered by a CMB photon during
its cosmic journey is about three arcminutes, the deflections
themselves are coherent over several degrees, which is	comparable
to the typical size of the acoustic features on the CMB. Thus
lensing causes coherent distortions of the hot and cold spots on
the CMB, and thereby broadens their size distribution. This leads
to redistribution of power among the acoustic scales in the CMB,
and shows up in the two-point statistics as a  smoothing of the
acoustic peaks. At smaller scales, where the primordial CMB is well
approximated by a {local} 
gradient, deflectors of small angular size produce
small-scale distortions in the CMB, thereby transferring power from
large scales in the CMB to the higher multipoles. Also, although the
primordial CMB can be safely assumed to be a Gaussian random field
\citep{2003ApJS..148..119K}, and the large scale lensing potential can
also be well approximated by  a Gaussian random field, the lensed
CMB--- being a reprocessing of one Gaussian random field by another---
is itself not Gaussian. The effect of lensing on the power spectrum
of the CMB is important enough that it should be
taken into account while deriving parameter constraints with future
higher resolution experiments. But what is even more interesting is that
the non-Gaussianity in the lensed CMB field should
enable us to extract information about the projected large scale
structure potential, and thereby constrain the late time evolution
of the Universe and Dark Energy properties. 
Therein lies the main motivation of studying this
effect in utmost detail.
Progress in this area has been slow.  Measurements
of the CMB precise enough to enable a detection of weak lensing
were not available in the pre-WMAP era. 
Also, picking out non-Gaussian signatures in the measured
CMB sky by itself is
extremely difficult, due to confusion from systematics, foregrounds,
and limited angular resolution. 

Rather than looking at signatures
of lensing only in the CMB, one can also measure to what
extent the deflection field estimated from the CMB correlates with
tracers of the large scale structure which contributed to the lensing.
It is easily realized that this approach is
powerful \citep{2000ApJ...540..605P} because many of the
systematics disappear upon cross-correlating data sets. This
approach was taken in recent years by \cite{2004PhRvD..70j3501H}
and \cite{2007PhRvD..76d3510S}, using WMAP 1-year and
3-year data respectively. The former work looked at the cross
correlation with SDSS luminous red galaxies (LRG), while the latter used the NRAO-VLA Sky
Survey (NVSS) radio sources as their large scale structure tracers.
As the lensing efficiency for the CMB is highest between redshifts
of one and four, higher redshift tracers should show  greater cross
correlation signal, which makes the NVSS radio sources better tracers
for such study; \cite{2007PhRvD..76d3510S}
report a $3.4\sigma$ detection. An independent analysis by \citet{2008arXiv0801.0644H} looking for this effect in the WMAP 3-year data in cross
correlation with SDSS LRG+QSO and NVSS sources find this signal at the $2.5\sigma$ level.  With these pioneering
efforts and with higher resolution CMB data from experiments
such as ACT,
PLANCK and the  South Pole Telescope (SPT)\footnote{http://spt.uchicago.edu}
on the horizon, we are entering an era where robust detection
and characterization of this effect will become a reality. Also,
with upcoming and proposed large scale structure projects
(LSST\footnote{\url{http://www.lsst.org/lsst\_home.shtml}},
SNAP\footnote{\url{http://snap.lbl.gov/}},
ADEPT\footnote{\url{http://universe.nasa.gov/program/probes/adept.html}},
DESTINY\footnote{\url{http://destiny.asu.edu/}}, etc.) there will
in future be many more datasets to cross-correlate with the CMB.

One of the immediate results of such cross-correlation
studies will be a measurement of the bias of the tracer
population. Because such cross correlations tie together early
universe physics from the CMB and late time evolution from
large scale structure, they will also be sensitive to Dark
Energy parameters \citep{2006ApJ...650L..13H} and neutrino
properties \citep{2006PhRvD..74l3002S,2006PhRvD..73d5021L},
and can potentially break several parameter degeneracies in
the primordial CMB (M.~Santolini,  S.~Das and D.~N.~Spergel, in
preparation). Combination of galaxy or cluster lensing of the CMB
with shear measurements from weak lensing of galaxies can also provide
important constraints on the geometry of the Universe \citep[S.~Das
and D.~ N.~Spergel, in prep;][]{2007arXiv0708.4391H}. Again, with
high enough precision of CMB data, it is possible to estimate,
using quadratic \citep{2003PhRvD..67h3002O}	or maximum
likelihood \citep{2003PhRvD..67d3001H} estimators, the deflection
field that caused the lensing. Such estimates can be turned into
strong constraints of the power spectrum of the projected lensing
potential \citep{2002ApJ...574..566H}, which is also sensitive
to the details of growth of structure. The estimated potential
from the lensed CMB alone, or the potential estimated from weak
lensing surveys \citep{2007arXiv0710.2538M}, can be also used to
significantly de-lens the CMB. This is particularly important in
the detection of primordial tensor modes via measurements of CMB
polarization. This is because (even though detection of the so-called
B modes in CMB polarization is hailed as the definitive indicator
of the presence of gravitational waves from the inflationary era)
these mode can be potentially contaminated by the conversion of
E-modes into B-modes via gravitational lensing. De-lensing provides
a way of cleaning these contaminating B-modes produced by lensing
and thereby probing the true gravitational wave signature.

In this paper, we describe a method for simulating the
gravitational lensing of the CMB temperature field on a large
area of the sky using a high resolution Tree-Particle-Mesh
\citep[TPM;][]{2000ApJS..128..561B,2003ApJS..145....1B} simulation of large
scale structure to produce the lensing potential. The reason for
considering a large area of the sky is twofold.	First,
the deflection field has most of its power
on large scales (the power spectrum of the deflection field peaks at $\ell\sim 50$ in the best-fit cosmological model), and much of the power redistribution in the
acoustic peaks of the CMB occurs via coupling of modes in the
CMB with these large coherent modes in the deflection field. A
large sky allows for several such modes to be realized. It is
estimated that a small (flat) sky simulation that misses these
modes would typically underestimate the lensing effect by about
$10\%$ in the acoustic regime, and more in the damping tail
\citep{2000PhRvD..62d3007H}. Second, one of the major goals of
simulations such as this is to produce mock observations for 
upcoming CMB experiments.  PLANCK is an all-sky experiment,
and many of the future CMB experiments (including ACT and SPT) will
observe relatively large patches of the sky. Therefore, simulating
CMB fields on a large area  of the sky is a necessity.  This
method fully takes into account the curvature of
the sky.  Although presented here for a polar cap like area, it
can be trivially extended to the full sky.

The value of a simulation as described here is multifaceted,
particularly
in the development of algorithms for detection and characterization
of the CMB lensing effect for a specific experiment. Since
each experiment has a unique scanning mode, beam pattern, area
coverage, and foregrounds, operations and optimizations 
performed on the data to extract the lensing information will
have to be tailored to the specific experiment.  A large-sky 
lensed CMB map acting as an input for a telescope simulator
provides the flexibility of exploring various observing strategies,
and also allows for superposition of known foregrounds.  Another
important aspect of this simulation is that the halos identified in
the large scale structure simulation can be populated with different
tracers of interest. Also, other signals, such as the Thermal
and Kinetic Sunyaev Zel'dovich	effects and {weak lensing of galaxies by large scale structure}, can be
simulated using the same large scale structure. This opens up the
possibility of studying the  cross-correlation of the CMB lensing
signal with various indicators of mass, and thereby predicting the
level of scientific impact that a specific combination of experiments
can have.

As noted in \cite{2005PhRvD..71h3008L}, the
exact simulation of the lensed CMB sky, which requires the
computation of spin spherical harmonics on an irregular grid
defined by the original positions of the photons on the CMB
surface, is  computationally expensive and requires robust
parallelization.
\cite{2005PhRvD..71h3008L} suggested an alternative in which
one would resample an unlensed CMB sky, generated with finer
pixelation, at these unlensed positions. This method was implemented in 
the publicly available LensPix\footnote{http://cosmologist.info/lenspix} code
that was based on \cite{2005PhRvD..71h3008L}.
However, producing a
high resolution lensed map requires 
a much higher resolution unlensed map, the generation of which
becomes computationally more expensive as resolution increases.
Here we put forward another alternative, in which we
do the resampling with a combination of fast spherical harmonic
transform on a regular grid followed by a high order polynomial
interpolation. This interpolation scheme has been adapted from
\cite{2004PhRvD..70j3501H}, and is called the Non-Isolatitude
Spherical Harmonic Transform (NISHT). This method is accurate as
well as fast, and does not require parallelization or production of
maps at a higher resolution. Another added advantage of this method
is that the same algorithm can be used to generate the gradient of
a scalar field on an irregular grid. Since the deflection field is a
gradient of the lensing potential, this opens up the possibility of
performing  a multiple plane ray tracing simulation. This is because
the rays, as they propagate from one plane to another, end up on
irregular grids, so the deflection fields on the subsequent planes
have to be evaluated on irregular grids.
At the time of the development of this project,  LensPix did not include an
interpolation scheme, and used the methods as described originally
in that paper. Concurrently with the completing of the current work,
an interpolation scheme \citep{232854}  different from the one described here has been added to that code. Another notable difference of our results with LensPix, is that while the latter uses a Gaussian Random realization of the deflection field, we have used a large scale structure simulation to produce the same, thereby including all higher order correlations due to non-linearities.  

The paper is laid out as follows. In \S\ref{Algorithm} we
explain the lensing algorithm,	describing the governing
equations in \S\ref{Equations} and their discretization
in \S\ref{Discretization}. Then we discuss the effective
lensing approach (\S\ref{Effective_Lensing}) as well as the
multiple plane ray tracing approach (\S\ref{Multiple_Plane}).
At the heart of the lensing algorithm lies the non-isolatitude
spherical harmonic transform algorithm adapted from
\cite{2004PhRvD..70j3501H}, which is reproduced in some detail for
completeness in \S\ref{Interpolation}. As discussed earlier,
we have employed a light cone $N$-body  simulation and adopted a
special polar cap like geometry for generating the lensing planes
(\S\ref{Plane_Generation}). For comparison of the simulated fields
with theoretical prediction, we compute the angular power spectra on
the polar cap window; in \S\ref{Angular_Power_Spectra} we describe
some of the subtleties involved in computing the power spectra.
We present our results in \S\ref{Results} and describe the tests
that we have performed in \S\ref{Tests}.
Conclusions are presented in \S\ref{Conclusions}.

\section{ The Lensing Algorithm}\label{Algorithm}
\subsection{ Basic Equations}\label{Equations}
We would like to note here that while the calculations for the
simulation described here has been done for a flat universe, our
approach is generalizable to non-flat geometries.

The deflection angle  of a light ray propagating through the space is
\beq
\label{deflection}
d\bm\alpha=-2\nabla_\perp \Psi d\eta  ,
\eeq
where is $d\bm\alpha$ is the deflection angle, $\Psi$ is the Newtonian potential, $\nabla_\perp$
denotes the spatial gradient on a plane perpendicular to light
propagation direction and $\eta$ is the radial comoving distance.
The transverse shift of the light ray position at $\eta$ due to a
deflection at $\eta'$ is given by
\beq
d\bmm x (\eta)=d_A(\eta-\eta') d\bm\alpha(\eta')  ,
\eeq
where $d_A(\eta)$ is the comoving angular diameter distance.

The final angular position $\bm \theta(\eta)=\bmm x(\eta)/d_A(\eta)$
is therefore given by
\beqn
\label{finalpos}
\nn\bm\theta(\eta)&=&\bm\theta(0)-\frac{2}{d_A(\eta)}\int_0^{\eta}
d\eta' d_A(\eta-\eta') \nabla_\perp \Psi\\
&=&\bm\theta(0)+\bm{\tilde{\alpha}}(\eta)  ,
\eeqn
where $\bm{\tilde\alpha}$ is the total effective deflection.

\subsection{\label{Discretization} Discretization}
We will now discretize the above equations by dividing the radial
interval between the observer and the source into $N$ concentric
shells each of comoving thickness $\Delta \eta$. We project the
matter in the $i$-th shell onto a spherical sheet at comoving
distance $\eta_i$ which is halfway between the the edges of the
shell ($i$ increases as one moves away from the observer).  Since we
shall be working in spherical coordinates it is advantageous to use
angular differential operators instead of spatial ones. We rewrite
\eqref{deflection} in terms of the angular gradient $\nabla_{\nhat}$
as
\beq
\label{deflection2}
d\bm\alpha=-\frac{2}{d_A(\eta)}\nabla_{\nhat} \Psi d\eta  .
\eeq
At the $j$-th shell at $\eta_j$, the deflection angle due to the
matter in the shell can be approximated by an integral of the
above:
\beqn
\bm\alpha^j&=&-\frac{2}{d_A(\eta_j)}\int_{\eta_j-\Delta\eta/2}^{\eta_j+\Delta\eta/2}
\nabla_{\nhat} \Psi(\tilde\eta\nhat;\tilde\eta) d\tilde \eta\\
&=&-\nabla_{\nhat} \phi^j(\nhat)  ,
\eeqn
where we have defined the 2-D potential on the sphere as
\beq
\phi^j(\nhat)=\frac{2}{d_A(\eta_j)}\int_{\eta_j-\Delta\eta/2}^{\eta_j+\Delta\eta/2}
\Psi(\tilde\eta\nhat;\tilde\eta)d\tilde\eta.
\eeq
Here, the notation $(\eta\nhat;\eta)$ signifies that the potential is evaluated at the conformal look-back time $\eta$, when the photon was at the position $\eta\nhat$.
The potential can be related to the mass overdensity in the shell
via  Poisson's equation, which reads
\beqn
\nabla^2_{\eta}\Psi&=&\frac{4\pi G}{c^2} 
\frac{\rho-\bar{\rho}}{\left( 1+z \right)^2}  ,
\eeqn
$\bar\rho$ being the mean matter density of the universe at 
redshift $z$. 
By integrating the above equation along the line of sight, one can arrive at a two dimensional version of the Poisson equation \citep{2003ApJ...592..699V}, 
\beqn
\label{2Dpoisson}
\nabla^2_{\nhat} \phi^j(\nhat)=\frac{8\pi G}{c^2} \frac{d_A(\eta_j)}{
(1+z_j)^{2}}\Delta_\Sigma^j(\nhat)
\eeqn
where the surface mass density
\beq
\label{def:Delta_Sigma}
\Delta_\Sig^j=\int_{\eta_j-\Delta\eta/2}^{\eta_j+\Delta\eta/2}(\rho-\bar{\rho})
d\tilde\eta  .
\eeq
Note that in going from the three dimensional to the two dimensional version, the term containing the radial derivatives of the Laplacian can be neglected \citep{2000ApJ...530..547J}. One can show that this term is small by expanding the potential $\Psi$ in Fourier modes $\bmm k$, with components $k_\parallel$ parallel to the line of sight and $k_\perp$ transverse to it.  Then, the ratio of the components of the line of sight integral in the parallel and transverse directions will be $\sim k_\parallel^2/k_\perp^2$. Due to cancellation  along the line of sight, only the modes with wavelengths comparable to the line of sight depth of each slice will survive the radial integral. These would be the modes with $k_\parallel \lesssim \frac{2\pi} {\Delta \eta}$. On the other hand, the transverse component gets most of its contribution from scales smaller than $\sim 100$ Mpc i.e. $\k_\perp \gg 2\pi/100 \sim 0.1$ Mpc$^{-1}$. Under the effective lensing approximation,  the projection is along the entire line of sight from zero redshift to the last scattering surface, $\Delta \eta \sim 10^4$ Mpc, giving $\k_\parallel \lesssim 10^{-3}$ Mpc$^{-1}$. Therefore, in this case the ratio of the radial and transverse components of the integral will be  $\sim k_\parallel^2/k_\perp^2 \ll 10^{-4}$. For a multiple plane case, we would typically employ $10$ lensing planes for which this ratio would be $\ll 10^{-2}$. The approximation will break down if we employ thin shells.  \par
Defining the field $K$ as
\beq
\label{def:K}
K^j(\nhat)=\frac{4\pi G}{c^2} \frac{d_A(\eta_j)}{
(1+z_j)^{2}}\Delta_\Sigma^j(\nhat)  ,
\eeq
\eqref{2Dpoisson} takes the form
\beq
\label{Phi_K}
\nabla^2_{\nhat} \phi^j(\nhat)=2 K^j(\nhat)  .
\eeq
It is convenient to define an angular surface mass density
$\Delta_\Sigma^\theta(\nhat)$ as the mass per steradian,
\beq
\Delta_\Sigma^{\theta j}(\nhat) =
\int_{\eta_j-\Delta\eta/2}^{\eta_j+\Delta\eta/2} (\rho-\bar{\rho})
\frac{d_A(\tilde\eta)^2}{(1+\tilde z)^3}  d\tilde\eta .
\eeq
The surface mass density defined in
\eqref{def:Delta_Sigma} is related to this through the relation
\beq
\label{def:sigma_theta}
\Delta_\Sigma=\Delta_\Sigma^{\theta}\frac{(1+z)^3}{d_A(\eta)^2}  .
\eeq
This implies the following form of \eqref{def:K},
\beq
\label{def:Kth}
K^j(\nhat)=\frac{4\pi G}{c^2} \frac{
(1+z_j)}{d_A(\eta_j)}\Delta_\Sigma^{\theta j}(\nhat)  .
\eeq
Equation (\ref{def:Kth}) is the key equation here. The quantity $K$
can be readily calculated once the mass density is radially projected
onto the spherical sheet. Expanding both sides of \eqref{Phi_K}
in spherical harmonics, one has the following relation between
the components:
\beq
\label{philm}
\phi_{\lm}=\frac{2}{l(l+1)}K_{\lm}  .
\eeq
It is interesting to note that the apparently divergent monopole $(l=0)$ modes in the lensing potential can be safely set to zero in all calculations, because a monopole term in the lensing potential does not contribute to the deflection field.
Being the transverse gradient of the potential, the deflection angle $\bm \alpha(\nhat)$ is a vector (spin 1) field defined on the sphere and can be synthesized from the spherical harmonic components of the potential in terms of vector spherical harmonics, as will be described in \S~\ref{Interpolation}. 

\subsection{\label{Effective_Lensing} Connection with effective
lensing quantities}

%\clearpage
\begin{figure}
\center
\includegraphics[scale=0.3]{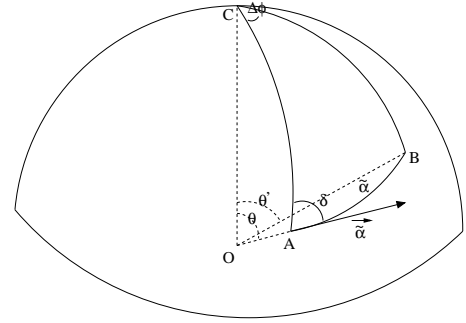}
\caption{\label{SphericalTriangle} Geometry illustrating the point
remapping used in the text}
\end{figure}
%\clearpage

In weak lensing calculations, one often takes an effective approach,
in which one approximates the effect of deflectors along the entire
line of sight by a projected potential or a convergence which is
computed along a fiducial undeflected ray (often referred to as
the Born approximation). One therefore defines an effective lensing
potential out to comoving distance $\eta_s$ as
\beq
\phi^{eff}(\nhat)=2\int_0^{\eta_s} d\eta
\frac{d_A(\eta_s-\eta)}{d_A(\eta)d_A(\eta_s) }\Psi(\eta \nhat;\eta)  .
\eeq
In terms of the projected potential, the effective deflection
(see Eq.~\ref{finalpos}) is given  by the angular gradient,
$\bm{\tilde{\alpha}}= - \nabla_{\nhat} \phi^{eff}$. An effective
convergence is also defined in a similar manner:
\beqn
\k(\nhat)&=&\frac{1}{2} \nabla_{\nhat}^2 \phi^{eff}(\nhat)\nn\\
&=& \int d\eta \frac{d_A(\eta_s-\eta)d_A(\eta)}{d_A(\eta_s)}
\nabla_{\perp}^2 \Psi(\eta \nhat;\eta)  .
\eeqn
In terms of the fields $\phi^j$ and $K^j$ defined on the multiple
planes, these quantities are immediately identified as the following
sums,
\beqn
\phi^{eff}(\nhat)\simeq \sum_j
\frac{d_A(\eta_s-\eta_j)}{d_A(\eta_s)}\phi^j(\nhat)  ,\\
\label{eq:kappa}
\kappa(\nhat)\simeq \sum_j \frac{d_A(\eta_s-\eta_j)}{d_A(\eta_s)} K^j(\nhat) .
\eeqn

Once $\kappa$ is obtained one can go through the analog of equation
(\ref{philm}) and take its transverse gradient to obtain the effective deflection
$\tilde{\bm{\alpha}}$. Using this, one can find the source position corresponding to the observed position $\theta(0)$:
\beq
\label{eq:effective_lensing}
\bm \theta_s=\bm \theta(0)+\tilde{\bm{\alpha}}.
\eeq
In \S\ref{Results}, we shall use this effective or single plane
approximation to lens the CMB.

Equation (\ref{eq:effective_lensing}) is to be interpreted in the
following manner \citep{2002PhRvD..66l7301C}. 
The effective deflection angle is a tangent
vector at the undeflected position of the ray. The  original
position of the ray on the source, or unlensed, plane is to be found
by moving along a geodesic on the sphere in the direction of the
tangent vector and covering a length $\tilde{\alpha}$ of an arc. The
correct remapping equations can be easily derived from identities of
spherical triangles \citep{2005PhRvD..71h3008L}. For completeness,
we give the derivation here.

In Fig.~\ref{SphericalTriangle}, let the initial and final position of
the ray in question be the points A$\equiv(\theta,\phi)$ and B$\equiv
(\theta^\prime,\phi+\Delta\phi)$, respectively. The North pole of
the sphere is indicated as C, so that the dihedral angle at A is
also the angle between $\tilde{\bm{\alpha}}$ and $-\etheta$, so that
\beq
\tilde{\bm{\alpha}}=-\tilde\alpha \cos\delta \etheta+ \tilde\alpha
\sin\delta \ephi  .
\eeq
Now, applying the spherical cosine rule to the triangle ABC, we have
\beq
\cos\theta^\prime=\cos\theta\;\cos\tilde\alpha+\sin\theta\sin\tilde\alpha\cos\delta  ,
\eeq
and applying the sine rule
\beq
\sin\Delta\phi=\sin\tilde\alpha \frac{\sin\delta}{\sin\theta^\prime}  .
\eeq
We use these equations to remap points on the CMB sky and on
the intermediate spherical shells in the multiple plane case,
as described below.
\subsection{\label{Multiple_Plane} Multiple plane ray tracing}
%\clearpage
\begin{figure}
\center
\includegraphics[scale=0.35]{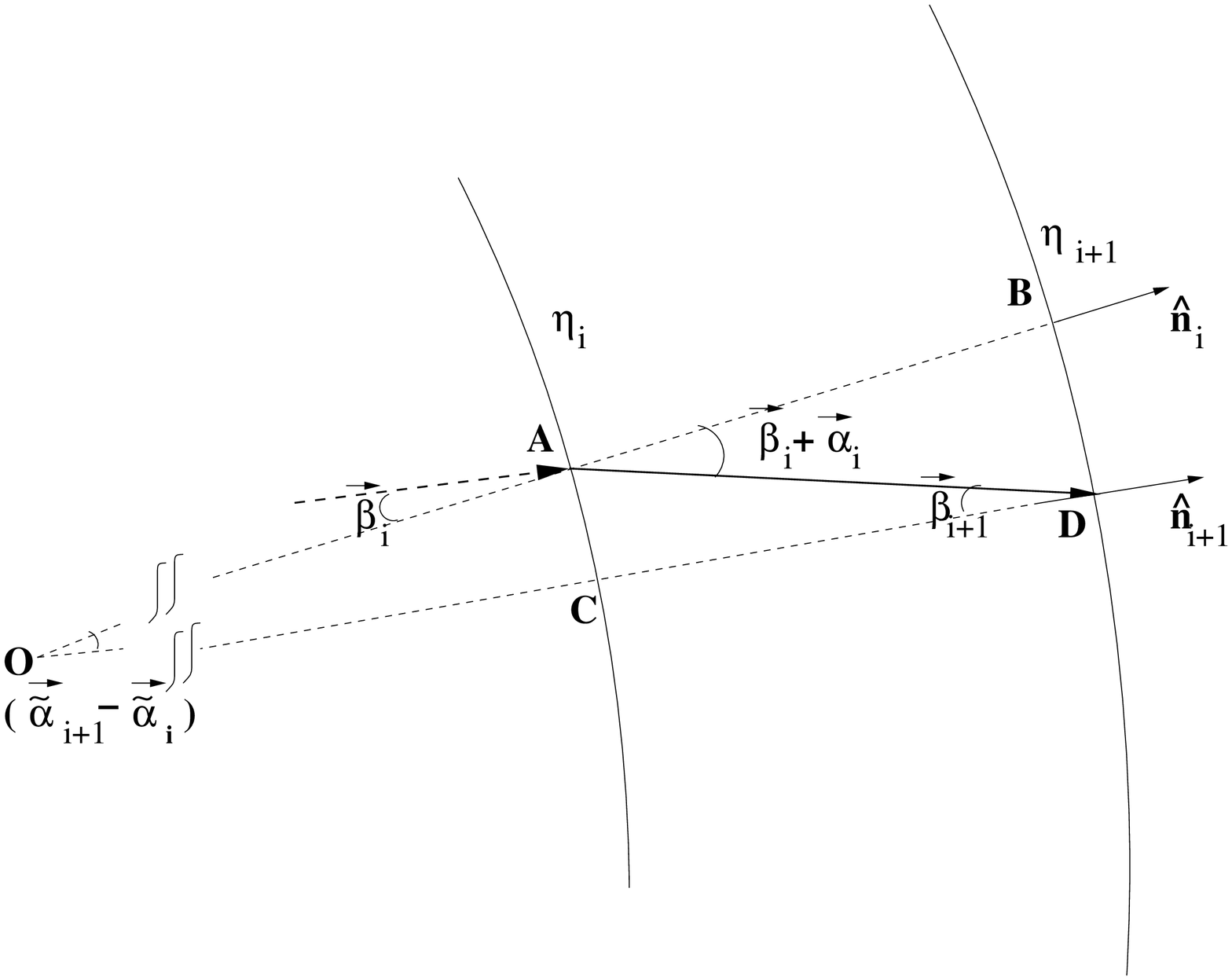}
\caption{\label{multiPlane} Geometry illustrating the multiple
plane ray tracing method.}
\end{figure}
%\clearpage

In the multiple plane case, we shoot ray outwards from the common
center of the spherical shell (i.e. the observer) and follow
their trajectories out to the CMB plane, thereby studying the
time reversed version of the actual phenomenon.  We assume all
intermediate deflections are small, as is really the case.
Here we describe how we keep track of a ray propagating
between multiple planes, as shown in Fig.~\ref{multiPlane}. We
assume a flat cosmology for this purpose.  At some intermediate stage
of the ray propagation, let a ray be incident on the $i$-th plane at
the point A, where it gets deflected  and reaches the $i$+1-th plane
at the point D. The ray incident at A will not in general lie on
the same plane as defined by the deflected ray $\vec{AD}$ and the
center O of the sphere, which we also consider as the plane of the
figure. Assuming that we know the incidence angle $\bm{\beta_i}$,
we can obtain  the additional angle of deflection $\bm\alpha_i$
due to the matter on plane $i$ and compute
the net deflection $\bm\alpha_i+\bm\beta_i$. {Let us denote by  $\tilde{\bm\alpha}_i$, the effective angle, by which a ray has to be remapped from its observed position $\bm\theta(0)$ to its current position $\bm\theta_i$ on plane i, so that
\beq
\bm\theta_i = \bm\theta(0) + \tilde{\bm\alpha}_i .  
\eeq 
Obviously, $\tilde{\bm\alpha}_1=0$ and $\bm\theta_1 = \bm\theta(0)$.
Therefore, the effective angle $(\tilde{\bm{\alpha}}_{i+1}-\tilde{\bm{\alpha}}_{i})$ by which the ray has to be remapped from
point B to point D on  the shell $i$+1 can be readily calculated from
two descriptions of the arc BD,
\beq
\label{effective_angle_i+1}
\eta_{i+1} (\tilde{\bm{\alpha}}_{i+1}-\tilde{\bm{\alpha}}_{i})=(\eta_{i+1}-\eta_i)
(\bm\alpha_i+\bm\beta_i)  .
\eeq
In order to repeat this process for the ($i$+2)-th shell, one needs
to know the value of the new incidence angle $\bm\beta_{i+1}$. We
now equate two ways of finding the length of the arc AC,
\beq
\eta_i (\tilde{\bm{\alpha}}_{i+1}-\tilde{\bm{\alpha}}_{i})=(\eta_{i+1}-\eta_i) \bm\beta_{i+1}.
\eeq
Substituting $(\tilde{\bm{\alpha}}_{i+1}-\tilde{\bm{\alpha}}_{i})$ from
\eqref{effective_angle_i+1}, 
\beq
\label{incidence_angle_i+1}
\bm\beta_{i+1}=\frac{\eta_i}{\eta_{i+1}} (\bm\alpha_i+\bm\beta_i).
\eeq
Since we shoot the rays radially on the first plane,
$\bm\beta_1=0$; therefore equations (\ref{effective_angle_i+1})
and (\ref{incidence_angle_i+1}) can be used to propagate the ray
back to the CMB surface, which we take to be the $(N+1)$-th plane, i.e. $\bm\theta_s = \bm\theta_{N+1}$. Although we only discuss results obtained with the effective or single plane approximation here, the multiple plane version is straightforward to perform and will be reported elsewhere. 
}
\subsection{\label{Interpolation} Interpolation on the sphere}
In practice we have used the
\healpix\footnote{\url{http://healpix.jpl.nasa.gov}} \citep{gorski-2005-622} scheme to
represent fields on the sphere. At various stages of the lensing
calculation, an accurate algorithm for interpolation on the sphere
becomes a necessity.  In the effective lensing approximation,
the original positions of the rays
will in general be off pixel centers.  This implies
that the lensed CMB field is essentially generated by sampling
the unlensed CMB surface at points which are usually not pixel
centers. Hence, obtaining the lensed CMB field is essentially an
interpolation operation. In the case of multiple lensing planes,
it is again obvious that (except for the first plane, on which
we can shoot rays at pixel centers by way of convenience) the
deflection field itself has to be evaluated at off-center points on
all subsequent planes. So, together with the interpolation of the
temperature map. we need to go between spin-0 and spin-1 fields on
an arbitrary grid. Therefore, one needs, in general, a spherical
harmonic transform algorithm that can deal with an irregular grid
on the sphere.

For this purpose, we adopt the Non-isolatitude
Spherical Harmonic Transform (NISHT) algorithm proposed by
\cite{2004PhRvD..70j3501H}; details of the algorithm can be found
in Appendix A of that paper.  Here we have reproduced the key
equations for clarity, and described the salient
features of the general algorithm with special attention to aspects
which are relevant for the current application. 

The basic operation for generating the lensed CMB maps can be broken
up into two steps:\\
\indent{\bf L1.} generating the deflection field on the sphere at
points where the rays land from the previous plane, and\\
\indent{\bf L2.}  sampling the unlensed CMB surface at the source-plane
positions of the rays to generate the lensed CMB field.

 Of course, in case of the effective lensing simulation, one can
 conveniently generate the deflection field at the pixel centers in
 step~{\bf L1} above. As step~{\bf L2} is a series of operations
 involving scalars and therefore conceptually simpler, we shall
 explain the NISHT algorithm in relation to this step. Step~{\bf
 L1}, which involves spin-1 fields on the sphere, is conceptually
 similar to the spin-0 case. 

The problem in step~{\bf L2} is that we know the CMB temperature
field $T(\nhat)$ on the \healpix  grid $\set{\nhat}$ as well as the
source-plane positions of the rays $\set{\nhat'}$ on the polar cap, and we
want to sample the CMB field at $\set{\nhat'}$. Suppose, by applying
the steps for spherical harmonic analysis (as will be described later),
we have the spherical harmonic components, $T_\lm$ of the temperature
field. Now, we need to synthesize the field using these $T_\lm$'s at
the points $\set{\nhat'}$. This operation can be formally written as
\beq
T(\nhat') = \sum_{\ell=0}^{\ell_{\mathrm{max}}}\sum_{m = -\ell}^{\ell}
T_\lm Y_\lm(\nhat'),
\eeq
where $\lmax$ is the Nyquist multipole and is set
by the resolution of the \healpix  grid as, $\lmax \simeq
\pi/\sqrt{\pixarea }$  (cf. equation \ref{eqn:pixarea}).
This synthesis operation can be split into the following four steps
(Eqns.~\ref{Latitude_Transform} through \ref{Interpolation_Weights}
are essentially reproduced for completeness from Appendix A of
\citealt{2004PhRvD..70j3501H}):
\begin{enumerate}
\item {\emph{ Coarse Grid Latitude Transform}}\\ 
As the first step,
we perform a transform in the latitude direction on an equally spaced
set of points, $(\theta =\pi \alpha/L, \phi = 0)$, where $\alpha$ is
an integer in the range $0\le\alpha\le	L$ and $L$ is 
a small integral multiple of some power of two
such that $L>\lmax$:
\beq
\label{Latitude_Transform}
T_m(\theta = \frac{\alpha}{L}\pi) = \sum_{\ell=\modu{m}}^{\lmax}
T_\lm Y_\lm(\theta = \frac\alpha L \pi, \phi = 0)  .
\eeq
The above calculation  involves $O(\lmax^2 L)$ operations.
\item \emph{Refinement of Latitude Grid}\\
  In this step we reduce the $\theta$ grid spacing from
  $\frac{\alpha}{L}\pi$ to  $\frac{\alpha}{L'}\pi$ where $L'>L$. We
  take advantage of the fact the sampling theorem can be applied
  to a linear combination of spherical harmonics which is band
  limited $(\ell \le \lmax)$ in the  multipole space, and hence
  can be written as a Fourier sum,
\beq
T_m(\theta) = \sum_{n=-\lmax}^{\lmax} C_{m,n} e^{in\theta}.
\eeq
We determine the coefficients $C_{m,n}$ via a fast Fourier
Transform (FFT) of length $2L$ and evaluate $T_m(\theta =
\frac{\alpha}{L'}\pi)$ using an inverse FFT of length $2L'$. This
step saves us the expensive generation of Associated Legendre
Polynomials on the finer grid. Each FFT  requires $O(\lmax L
\log(L))$ operations.
\item\emph{ Projection onto  Equicylindrical Grid}\\
Next, we perform the standard SHT step of taking an FFT in the
longitudinal direction to generate $T(\theta = \frac{\alpha}{L'}\pi,
\phi = \frac \gamma {L'} \pi)$,
\beq
T(\theta = \frac{\alpha}{L'}\pi, \phi = \frac \gamma {L'} \pi) =
\sum_{m = -\lmax}^{ m  =\lmax} T_m(\theta) e^{i m \phi}.
\eeq
 After this step we have  synthesized the map on an Equicylindrical
 projection (ECP) grid. The operation count for this step is $
 O(L'^2 \log  L')$ and the total operation count including this step
 is $O(\lmax^3)$.
\item \emph{Interpolation onto the final grid}\\
In the final step, given a required position $\nhat'$, we find the
nearest grid point in the ECP grid and determine the fractional
offset, $(\delta \alpha, \delta \gamma)$ between the two points,
\beq
\alpha + \delta \alpha = L' \frac \theta \pi; \: \gamma +\delta
\gamma = L' \frac \phi \pi.
\eeq
Then we perform a two dimensional polynomial interpolation using
$(2K)^2$ points around the nearest grid point, obtaining the value
at the required point as
\beq
T\simeq \sum_{\mu=-K+1}^{K} w_{\mu}(\delta_\alpha) \sum_{\nu=-K+1}^{K}
w_\nu(\delta_\gamma) T(\frac{\alpha+\mu}{L'}\pi, \frac {\gamma+\nu} {L'} \pi),
\eeq
with the weights computed using Lagrange's interpolation formula,
\beq
\label{Interpolation_Weights}
w_\rho(\delta) =
\frac{(-1)^{K-\rho}}{(K-\rho)!(K-1+\rho)!(\delta-\rho)}
\prod_{\sigma=-K+1}^{K} (\delta - \sigma).
\eeq
\end{enumerate}
{The inverse of the synthesis operation described above is the analysis operation, in which the spherical harmonic coefficients of a map defined on an irregular grid is needed.  This  can be thought of as the transpose of the above operations applied in reverse, and hence can be accomplished in an equal number of steps.}

The above algorithm can be easily extended to deal with vector
and tensor fields on the sphere. For a vector (spin 1) field,
the natural basis of expansion are the vector spherical harmonics,
\beqn
\nn \bmm Y_\lm^{V} &=& \frac{1}{\sqrt{\ell(\ell+1)}}\nabla Y_\lm\\
\bmm Y_\lm^{A} &=& \frac{1}{\sqrt{\ell(\ell+1)}} \nhat\times
\nabla Y_\lm,
\eeqn
where the superscripts $V$ and $A$ represent the ``vector-like''
and the ``axial-vector-like'' components, respectively.
In terms of these a  vector field $\bmm v(\nhat)$ can be expanded as,
\beq
\bmm v(\nhat)  =\sum_{\ell=0}^{\ell_{\mathrm{max}}}\sum_{m = -\ell}^{\ell} V_\lm \bmm Y_\lm^{V}(\nhat) + A_\lm
\bmm Y_\lm^{A}(\nhat).
\eeq
Therefore, given the ($V_\lm,A_\lm$) components one can go through
the analogs of the above steps for the scalar field synthesis. In
fact, to accomplish step {\bf L1} of the lensing algorithm, we go
from the convergence field $K$ to the deflection field,
\beqn
\nn \bm \alpha = - \nabla \phi &=& -\sum_{\ell=0}^{\ell_{\mathrm{max}}}\sum_{m = -\ell}^{\ell}  \phi_\lm \nabla
Y_\lm \\
\nn & = & - \sum_{\ell=0}^{\ell_{\mathrm{max}}}\sum_{m = -\ell}^{\ell}\frac{2}{\ell(\ell+1)}K_\lm \nabla Y_\lm\\
& = &  - \sum_{\ell=0}^{\ell_{\mathrm{max}}}\sum_{m = -\ell}^{\ell} \frac{2}{\sqrt{\ell(\ell+1)}} K_\lm \bmm Y_\lm^{V}.
\eeqn
Therefore, we go from the $K$ field on the polar cap
to the spherical harmonic components $K_\lm$ using the
analysis algorithm for scalar fields; then we divide the result
by $\sqrt{\ell(\ell+1)}/2$. This defines the vector field harmonic
components as $(V_\lm,A_\lm) = (-2 K_\lm/\sqrt{\ell(\ell+1)},0)$
from which we synthesize the deflection field at the required
points. 

The accuracy of the interpolation can be controlled by two
parameters: the rate at which  the finer grid oversamples the field
i.e. the ratio $L' / \lmax$, and the order of the polynomial $K$
used for the interpolation. Increasing either of these increases the
accuracy. In this paper we have used $L' = 4 \lmax$ and $K=10$,
which yields a fractional interpolation accuracy per Fourier mode
of $\sim 10^{-9}$.

\section{Generation of the lensing planes}\label{Plane_Generation}
{
An $N$-body dark matter simulation was performed to generate the
large scale structure; this same simulation has been discussed 
in \cite{2007ApJ...664..149S} and \cite{2007ApJ...663..139B},
so we refer the reader to these papers for more details.
Briefly, a spatially flat $\Lambda$CDM
cosmology was used, with a total matter density parameter $\Om=0.26$
and vacuum energy density
$\OL=0.74$. The scalar spectral index of the primordial power
spectrum was set to $n_s=0.95$ and the linear amplitude
normalized to $\sigma_8=0.77$. The present day value of the Hubble
parameter  $H_0=72 \: \hunit$.
A periodic box of size
$L=1000 \hmpc$ was used with $N=1024^3$ particles; therefore the
particle mass was $m_p=6.72\E{10} h^{-1}\msol$. The cubic spline softening
length was set to 16.28 $\hmpc$.
}
 
\subsection{\label{Projection} From the box to the sphere}
\begin{figure}[h]
\center
\includegraphics[angle=0,scale=0.3]{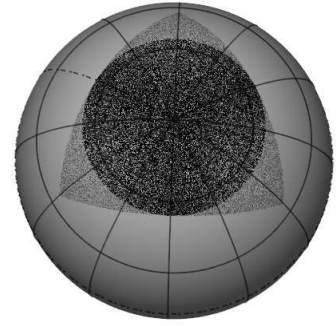}
\caption[CMBVIEW]{ \label{fig:polar cap} Illustration of the
Polar Cap geometry. The figure shows a 3-D rendering of the sphere
using the CMBVIEW\footnotemark  software, looking down towards the
North Pole. The lightly shaded triangular region correspond to the
positions of the particles in the octant from the $N$-body simulation
box at a typical time step.
The darker dots define what we call the Polar Cap
in the text. The surface mass density of the pixels
inside this Polar Cap region are saved in shells out to $z$=4. }
\end{figure}

{
We create the lensing planes on-the-fly from the $N$-body simulation. 
At each large time step (set by a Courant condition such that no
particle moves more than $\sim 122h^{-1}$kpc in this time)
the positions and velocities of the particles in a thin shell
are saved.  The mean radius of the shell is the comoving
distance to the redshift at that time, and the width 
(a few $\hmpc$) corresponds to the time step.  
Each shell is centered on the origin of the
simulation and covers one octant of the sky ($x,y,z>0$).
Note that for shells with radii greater than the simulation box
size, periodic copies of the box are used.  Thus a given structure
will appear more than once in the full light cone, albeit at 
different times and viewed from different angles.  }
We then Euler rotate the
coordinate axes so that the new $z$-axis passes through the centroid
of the octant. This is done to make the centroid correspond to the
North Pole on the \healpix  sphere.  We use the \healpix  routine
{\tt vec2pix} to find the pixel that contains the particle's
position on the sky.
We then place the mass of the
particle into that pixel by assigning to it the surface mass density
$\Sigma_p =m_p/\pixarea$, where \pixarea	is the area of a
pixel in steradians (cf. equation \ref{eqn:pixarea}).  
Thus, if $n$ particles fall inside the beam
defined by a pixel, then the pixel ends up having a surface mass
density of $n\Sigma_p$.  To simplify the geometry, we save only those
particles which fall inside a Polar Cap like region defined by the
disc of maximum radius that can be cut out from the octant (see
Fig.~\ref{fig:polar cap}). 

%\clearpage
%\clearpage
\footnotetext{http://www.jportsmouth.com/code/CMBview/cmbview.html}
By the end of the
run, $449$ such planes were produced from the simulation,
spanning $z=\maxz$ to $z=0$.
As these are far too many planes for the purpose of lensing,
we reduce them into $\sim 50$ planes by dividing up the original
planes into roughly equal comoving distance bins and adding up
the surface mass density pixel by pixel for all planes that fall
inside a bin to yield a single plane per bin. Hereafter, we shall
refer to the original planes from the TPM run as the TPM-planes
and the small number of planes constructed by projecting them as
the lensing-planes.
  
The angular radius of the Polar Cap is given by
$\theta_\mathrm{cap}=\arccos(2/\sqrt{6})$, and the solid angle
subtended by it is $\caparea=2\pi (1-\cos\theta_\mathrm{cap})=1.981
\:  \steradian =3785 \:\mathrm{sq-deg}$. Due to pixelation, the true
total area $\tildecaparea$
of the \ncap pixels that make up the Polar Cap is not
exactly equal to $\caparea$, but rather
\beq
\tildecaparea=\ncap \pixarea     .
\eeq
We will denote the surface mass density in pixel $p$ as $\sigma_p$
which has units of mass per steradian.

In \healpix, the resolution is controlled by the parameter NSIDE,
which determines  the number, $N_\mathrm{pix}$ of equal area pixels
into which the entire sphere is pixelated, through the relation 
$N_{\mathrm{pix}}=12 \times\mathrm{NSIDE}^2$,
so that the area of each pixel becomes,
\beq
\label{eqn:pixarea}
\pixarea=\frac{4\pi}{N_{\mathrm{pix}}}\:\: \mathrm{steradians}.
\eeq
The angular resolution is often expressed through the number
$\theta_\mathrm{res}=\sqrt{\pixarea}$   . It is also useful to define the fraction area of the sphere covered by the polar cap as,
\beq
\fsky = \frac{\tildecaparea}{4\pi} .
\eeq 

For results presented in this paper the resolution parameter NSIDE
was set to \nsidenow,  which corresponds to an angular resolution
of \angresnow.

\subsection{\label{Sigma_to_Kappa} From surface density to
convergence}
To construct the quantities required for lensing, we first convert
the surface mass density maps into surface over-density maps $\Delta
\Sig^{\theta}$  as defined in \eqref{def:sigma_theta}. It is straightforward to obtain the  $K$-maps defined in \eqref{def:Kth}
from the above map. Finally, \eqref{eq:kappa} is used to
obtain the effective convergence map on the Polar Cap. It is evident
that the convergence map constructed out of the simulated lensing
planes in this way will only contain the contribution from large
scale structure up to the redshift of the farthest lensing plane
($z =  4.05$). However, to accurately lens the CMB we need to add
in the contribution from higher redshifts up to the last scattering
surface. We do this by generating a Gaussian random field from a
theoretical power spectrum of the convergence between $z =  4.05$
and $z = z_{\mathrm{CMB}}$, {computed from the matter power spectrum obtained using CAMB, 
and adding it onto the convergence map from the TPM simulation}.
\subsection{The unlensed CMB map}
We used the \texttt{synfast} facility in \healpix  to generate
the unlensed CMB map. This takes as an input a theoretical unlensed
power spectrum and synthesizes a Gaussian-random realization
of the unlensed CMB field. For computing the theoretical power
spectrum we have used the publicly available Boltzmann transfer
code CAMB\footnote{\url{http://www.camb.info}}, with the same set
of cosmological parameters as used for the large scale structure
simulation.
\section{Measuring Angular Power
Spectra}\label{Angular_Power_Spectra}
At several stages we compute the power spectra of the maps to
compare with theory. For example, to verify that we have created the
convergence map correctly, the angular power spectrum of the $\kappa$
map is computed and compared to the theory. Also, we do the same for
the lensed map on the polar cap. We use the {\tt map2alm} facility
of \healpix  to perform a spherical harmonic decomposition of a
map $T(\nhat)$ on the Polar Cap. The resulting spherical harmonic
components, i.e.  $\tilde T^{\mathrm{pix}}_{\lm}$'s, are then combined to obtain
the pseudo-power spectrum,
\beq
\label{cl-tilde-pix}
\tilde\cl^{\mathrm{pix}}=\frac{1}{2\ell+1}\sum_{m = -\ell}^{\ell}
\modu{\tilde T^{\mathrm{pix}}_{\ell m}}^2 .
\eeq
There are two effects that need to be taken into account before
comparing the above result with theory, namely the finite pixel size,
signified by the superscript, $\mathrm{``pix''}$ and the incomplete
sky coverage, represented by the tilde.

{To simplify the following discussion of pixelation effects, for the moment we shall ignore the effect of the incomplete sky 
coverage. Also, we shall use the shorthand notation $\Sigma_\lm$ to denote the sum $ \sum_{\ell=0}^{\infty} \sum_{m = -\ell}^{\ell}$.}
  Due to the finite pixel size, a field realized on the HEALPix sphere
 is a smoothed version of the true underlying field, i.e. the value
 of the field in pixel $i$ is given by
\beq
T^{\mathrm{pix}(i)} = \int d^2\nhat w^{(i)}(\nhat) T(\nhat)  ,
\eeq
where $w^{(i)}$ is the window function of the $i$-th pixel as is
given by
\beq
w^{(i)}(\nhat) =\cases{\Omega^{-1}_{\mathrm{pix}}&inside pixel $i$,\cr
 0&elsewhere.}
\eeq
Expanding the true field $T$ in terms of spherical harmonics as
\beqn
\nn T(\nhat)  = \sum_{\lm} T_{\lm} Y_{\lm}(\nhat),
\eeqn
we have
\beq
\label{pixelized_field_lms}
T^{\mathrm{pix}(i)} = \sum_{\lm} w^{(i)}_{\lm} T_{\lm},
\eeq
where
\beq
\label{wlm}
w^{(i)}_\lm = \int d^2\nhat w^{(i)}(\nhat) Y_\lm(\nhat)
\eeq
is the spherical harmonic transform of the pixel window function. In
the \healpix  scheme, due to the azimuthal  variation of the pixel
shapes over the sky, especially in the polar cap area, a complete
analysis would require the computation of these coefficients for each
and every pixel. However, even for a moderate NSIDE, this calculation
becomes computationally unfeasible. Therefore, it is customary to
ignore the azimuthal variation and rewrite \eqref{wlm} as
\beq
\label{wlm_azimuth}
w_\lm^{(i)} = w_\ell^{(i)} Y_\lm(\nhat_i),
\eeq
 where one defines an azimuthally averaged window function
\beq w_\ell^{(i)} = \frac{4\pi}{(2\ell+1)} \left[
\sum_{m=-\ell}^{\ell} \modu{w_\lm}^2 \right]^{1/2}.
\eeq
From equations (\ref{wlm_azimuth}) and (\ref{pixelized_field_lms})
it immediately follows that the estimate of the power spectrum of
the pixelated field is given by
\beq
\cl^{\mathrm{pix}} = w_\ell^2 \av{T_\lm T_\lm^*} = w_\ell^2
\cl
\eeq
where one defines the pixel averaged window function,
\beq
w_\ell = \left(\frac1{N_{\mathrm{pix}}} \sum_{i=0}^{N_{\mathrm{pix}-1}}
(w_\ell^{(i)})^2\right)^{1/2}.
\eeq
This function is available for $\ell<4\times $NSIDE in the \healpix  distribution. We take
out the effect of the pixel window by dividing the computed power
spectrum by the square of the above function. {Coming back to the case at hand, where we
have both pixel and incomplete sky effects, we recover the power spectrum $\tilde\cl$ after correcting for the pixel window function in this manner.}

The second and more important effect that one needs to take into
account results from the fact that our field is defined only inside
the polar cap. This is equivalent to multiplying a full sky map
with a mask which  has value unity inside the polar cap and zero
outside. As is well known, such a mask leads to a coupling between various
multipoles, leading to a power spectrum which is  biased away from the
true value. As this effect tends to move power across multipoles,
the problem is	more acute for highly colored power spectra like
the CMB. 

Let us denote the effective all-sky mask with $W$, where
\beq
\label{eq:polcapwindow}
W(\nhat) =\cases{1 &	      {inside the polar cap,}\cr
0 &  elsewhere.}
\eeq
The spherical harmonic components of the masked field is therefore
given by
\beqn
\tilde T_\lm &=& \int d^2\nhat T(\nhat) W(\nhat) Y^*_\lm(\nhat)\\
& = & \sum_{\ell'm'} T_\lmprime \int  d^2\nhat Y_{\ell' m'} (\nhat)
W(\nhat)  Y^*_\lm(\nhat)
\eeqn
and the measured power spectrum by \citep[see for
example][]{2002ApJ...567....2H}
\beqn
\label{pseudocl}
\nn{\tilde C_{\ell_1}} &=& \frac1{(2\ell_1+1)}\sum_{m_1 =
-\ell_1}^{\ell_1} \av{\tilde T_\lmone \tilde T_\lmone^*}\\
& = & \sum_{\ltwo} M_{\lone\ltwo} C_\ltwo
\eeqn
where $C_\ltwo$ is the true power spectrum and $M$ is the mode
coupling matrix given by
\beq
\label{Ml1l2}
M_{\lone\ltwo} = \frac{(2\ltwo+1)}{4\pi} \sum_{\lthree} (2\lthree+1)
{\cal W}_{\lthree} \left(
\begin{array}{ccc}
0 & 0 & 0\\
\lone & \ltwo & \lthree
\end{array}
 \right)^2,
\eeq
with the power spectrum of the mask defined as 
\beq
{\cal W}_{\ell}  = \frac1{2\ell+1} \sum_m \modu{W_\lm}^2,
\eeq$W_\lm$ being the spherical harmonic components of the mask
$W(\nhat)$.

For a polar cap with angular radius $\Theta$, this function is
analytically known to be \citep{Dahlen:2007sv}
\beq
\label{w_l_cap}
{\cal W}_{\ell}^{\mathrm{cap}} = \frac \pi{(2\ell+1)^2} \left[P_{\ell -
1}(\cos \Theta) - P_{\ell+1}( \cos\Theta\right)]^2.
\eeq
where $P_\ell$	is a  Legendre Polynomial of order $\ell$ and
$P_{-1}(\mu) = 1. $ 

%\clearpage
\begin{figure}
  \center
  \includegraphics[scale=0.4]{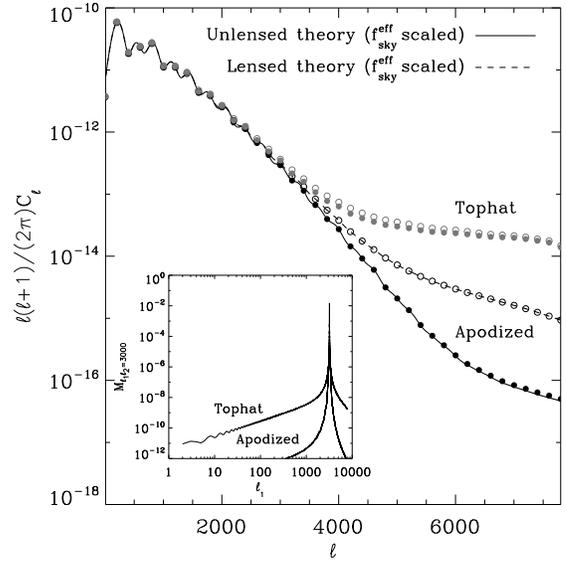}
  \caption{\label{ModeCoupling} Effect of apodization of the
  window function. The continuous line is the unlensed CMB power
  spectrum and the dashed line is the lensed one. Both have been
  scaled by  $\fsky^{\mathrm{eff}}$, the effective fractional sky
  coverage (see text). The gray filled and open circles labeled
  ``Tophat'', represent, respectively, the theoretical unlensed
  and lensed power spectra convolved with a window function that
  is unity inside the polar cap and zero outside. The black filled
  and open circles represent the same quantities, but in the case
  of a window which is apodized at the edge of the polar cap, as
  discussed in the text. Aliasing of power to higher multipoles
  due to mode coupling is significantly reduced in the latter
  case. We use the apodized window to mask the polar cap maps for
  computing various power spectra, and use the corresponding theory
  power spectrum convolved  with the same window for comparing
  our results with theory. (\emph{Inset:} The mode coupling matrix
  $M_{\lone\ltwo}$ as  a function of $\ell_1$, for $\ell_2=3000$,
  showing the reduction in the power in off-diagonal elements as
  a result of apodization.)  }
\end{figure}
%\clearpage

The  window function in \eqref{eq:polcapwindow} corresponding to the
polar cap is a ``tophat'' in the sense that it abruptly falls to zero
at the edge. The power spectrum (\eqref{w_l_cap}) of this window
has an oscillatory behavior showing a lot of power over a large range
of multipoles, an effect sometimes called \emph{ringing}.  Ringing
causes the mode coupling matrix, $M_{\ell \ell^\prime}$  to develop
large off-diagonal terms, as illustrated in Fig.~\ref{ModeCoupling},
and consequently the value of the measured power spectrum at any
multipole (\eqref{pseudocl})  has non-trivial contributions
from many neighboring multipoles. This causes the measured power
spectrum to be	biased, and its effect is particularly evident for
power spectra with a sharp fall-off such as the CMB. As is evident from
Fig.~\ref{ModeCoupling},  the effect of mode coupling due to the
polar cap becomes a serious problem for the lensed and unlensed
power spectra starting at moderately low multipoles ($\ell \sim
2000$). Although in principle one could compare the measured power
spectrum with a theoretical power spectrum which has been convolved
with the same mode coupling matrix, the effect is so strong in this
case that the lensed and unlensed spectra almost overlap each other.
This problem can be mitigated in principle by inverting a binned
version of the mode coupling matrix and thereby decorrelating
the power spectra. However an easier and less
computationally expensive solution can be achieved
in the following manner. 

The off diagonal terms of the mode coupling matrix can be reduced
significantly by apodizing the polar cap window function. Parenthetically,
 we note that there exists a general method of generating tapers on a cut-sky map, so as to minimize the effect of  mode coupling. This is referred to as the multi-taper method \citep[S.~Das, A.~Hajian and D.~N.~Spergel, 2007, in preparation;][]{Dahlen:2007sv}. However, for our purpose, it suffices to define a simpler apodizing window as
\beq
\label{polcapwindow}
W(\nhat) =
\cases{
1 &	      {for } $\theta<\theta_0<\capangle$\cr
\sin(\frac\pi2
\frac{\theta_{\mathrm{cap}}-\theta}{\theta_{\mathrm{cap}}-\theta_0}) & {for } $ \theta_0<\theta<\capangle $\cr
0 &  {for } $\theta > \capangle $  .}
\eeq
The power spectrum of this window can be easily computed using
\healpix, and thus the mode coupling matrix can be readily
generated using \eqref{Ml1l2}. {We found that an apodization window with $(\theta_{\mathrm{cap}}-\theta_0) \simeq 1.2$ degree, corresponding to $\sim 80$ pixels, works extremely well without eating into too much of the map.}
A section of the mode coupling matrix and the corresponding convolved power spectrum are displayed
in Fig.~\ref{ModeCoupling}. This figure shows that the power
spectrum convolved with the apodized window function has negligible
mode coupling.  Parenthetically, 
it is interesting to note that
simply scaling the theory power spectrum by the fraction of the
sky covered, $\fsky$, seems to do a good job in mimicking the effect
of the partial sky coverage, at least for the lower multipoles. In
fact, this approximation is an exact result for a white power
spectrum. However, when the window is apodized, the effective area of
coverage, $\fsky^{\mathrm{eff}}= \int W^2(\nhat) d^2\nhat /4\pi$, goes down a little (by $\sim 2.5\%$
for our apodization). We use  $\fsky^{\mathrm{eff}}$ scaled theory
power spectra only in some plots in this paper. For the analysis, we
perform the full mode coupling calculation. Therefore, when comparing
the power spectrum of some quantity defined on the polar cap with
theoretical predictions, we first multiply the map by the apodized
window and compare the resulting power spectrum with the theoretical
power spectrum	mode-coupled through the same weighting window.

\section{Results}\label{Results}
\begin{figure}
\center
\includegraphics[scale=0.5]{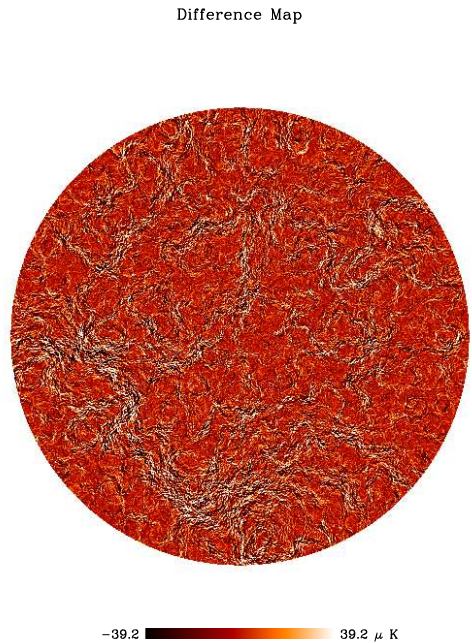}
\caption{\label{Difference_Map} The  Polar Cap map obtained after subtracting
the unlensed CMB map from the lensed CMB map. {To enhance the contrast, we have remapped the color scale to the range $(-2\sigma,2\sigma)$, $\sigma$ being the standard deviation of the map.} }
\end{figure}

We illustrate the algorithm with an effective lensing simulation
at the \healpix  resolution of $NSIDE=4096$. Since some rays end
up outside the polar cap after lensing, we have actually used
an unlensed CMB realization (using the {\tt synfast} facility
of \healpix) on an area larger than our fiducial polar cap to
accommodate those rays. As the gradient of the lensing potential is
ill defined at the edge of the polar cap, we ignore a ring of pixels
near the edge of the lensed map for all subsequent analyses. It is
particularly instructive to look at the difference of the lensed
and the unlensed maps, as shown in Fig.~\ref{Difference_Map}, as
it shows the large scale correlations imprinted on the CMB due to
the large scale modes in the deflection field. 

\begin{figure}
\center
\includegraphics[scale=0.37]{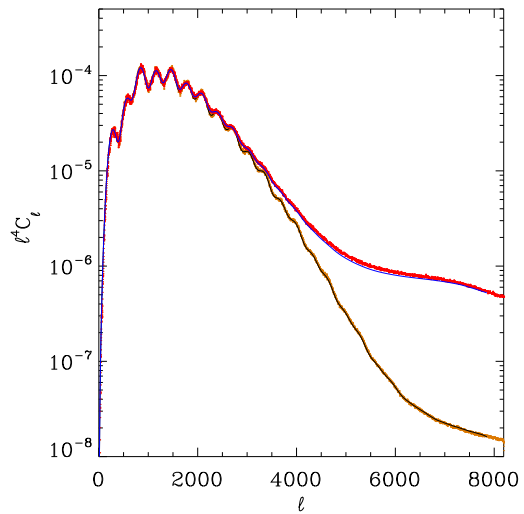}
\caption{\label{LensedCMBcls} The lensed and unlensed CMB
angular power spectra obtained from the simulation compared
with the theoretical models. The red and orange dots represent,
respectively, the lensed and unlensed angular power spectra
obtained from the polar cap using the methods described in \S
\ref{Angular_Power_Spectra}. The  solid black curve signifies the
theoretical unlensed power spectrum taking into account the
mode coupling due to the apodized polar cap window function. The
blue solid curve represents the same for the lensed power spectrum.}
\end{figure}

\begin{figure}
\center
\includegraphics[scale=0.37]{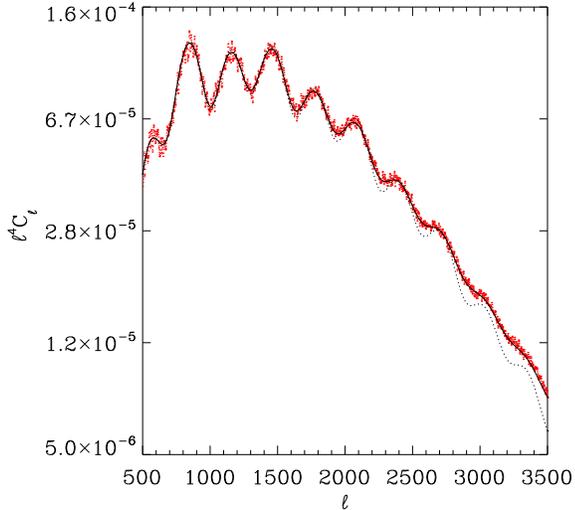}
\caption{\label{LensedCMBclsZOOM} Lensed CMB angular power spectrum
in the multipole range $500 <\ell<3500$ obtained from the simulation
compared with the theoretical model. The red dots represent
the lensed angular power spectrum obtained from the polar cap using
the methods described in \S \ref{Angular_Power_Spectra}. The solid black  
curve signifies the theoretical lensed CMB power spectrum taking
into account the mode coupling due to the apodized polar cap window
function. The dotted black curve represents the same for the theoretical
unlensed power spectrum and is shown here for contrast to the
lensed case.}
\end{figure}

We compute the angular power spectrum of the lensed and unlensed
CMB maps using an apodized weighting scheme as discussed in
\S\ref{Angular_Power_Spectra}. The resulting power spectra
are displayed in Fig.~\ref{LensedCMBcls} for the entire range
of multipoles analyzed, and are compared with the mode-coupled
theoretical power spectra.  The theoretical lensed CMB power
spectrum used for the calculation was generated with
the CAMB code, using the all-sky correlation function technique
\citep{2005PhRvD..71j3010C} and including nonlinear corrections
to the matter power spectrum.  In Fig.~\ref{LensedCMBclsZOOM},
we show a zoomed-in version of the lensed power spectrum, in the
multipole range $500 <\ell<3500$. From visual inspection of these
plots it is evident that the simulation does a good job
in reproducing the theoretically expected lensed power spectrum,
at least in the range of multipoles over which the computation
of the theoretical power spectrum is robust.  We defer a detailed
comparison of the simulation to the theory to
\S\ref{Tests_for_lensed_CMB}.
%\clearpage

%\clearpage

\section{Tests}\label{Tests}
\subsection{Tests for the mass sheets}

\begin{figure}
\center
\includegraphics[scale=0.37]{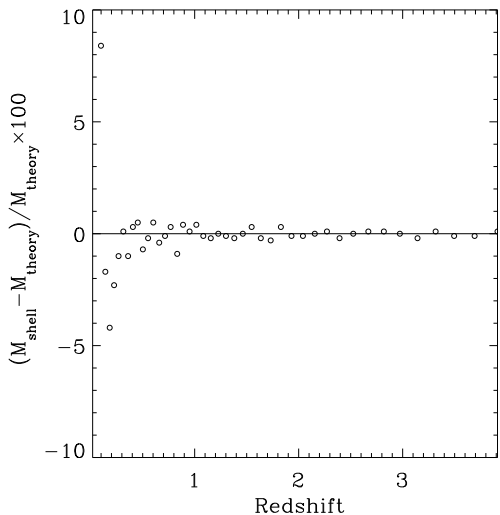}
\caption{\label{fig:binnedmasscompare}The mass in the lensing-slices
compared with that expected from theory.}
\end{figure}

In this section
we perform some sanity checks to ensure that the projection from
the simulation box onto the Polar Cap has been properly performed.
We first test that the total mass in each slice is equal to the
theoretical mass expected from the mean cosmology, the later being
given by
\beq
\mslice^{\mathrm{theory}}=\Om \rcr \bar \eta^2 \tildecaparea
\Delta \eta
\eeq
where $\Delta\eta$ is the comoving thickness of the slice at a
comoving distance $\bar\eta$. We compare this quantity with
\beq
\mslice=\sum_{i=1}^{\ncap}\sigma_{i}\pixarea
\eeq
which is the total mass on the plane from the simulation. 
The percentage difference between the two is depicted  
in Fig.~\ref{fig:binnedmasscompare} 
for the lensing-planes. Notice that the  agreement
is good to within 0.5\% for the high redshift planes, in which the
solid angle $\caparea$ corresponds to a large comoving area. For
low redshifts there are large variations due to the fact that
matter is highly clustered and $\caparea$ corresponds to a small
comoving area. These fluctuations at low redshift represent the
chance inclusion or exclusion of large dark matter halos within
the light cone.
%\clearpage
%\clearpage

\begin{figure}[h]
\center
\includegraphics[scale=0.37]{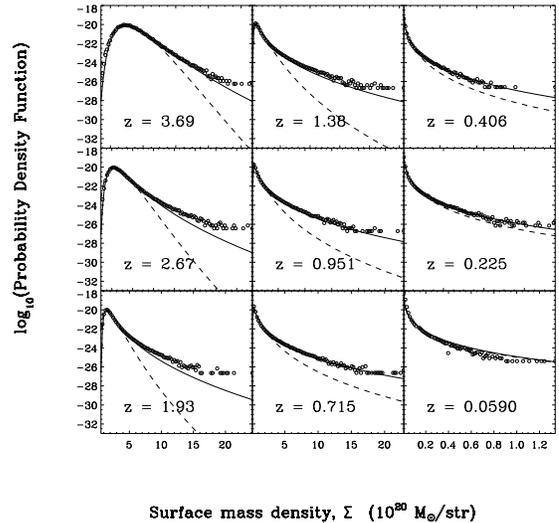}
\caption{\label{fig:pdfcompare} The probability density function
(PDF) of the surface mass density in the lensing-planes (circles) compared
with the lognormal (dashed line)  and the \cite{2006ApJ...645....1D}
model (solid line).  }
\end{figure}

Next, we make sure that the probability density function (PDF)
of the surface mass density is well behaved for each plane, and
is well modeled by analytic PDFs such as the  lognormal \citep{2001ApJ...561...22K,2002ApJ...571..638T} or the
model proposed by \cite{2006ApJ...645....1D}.  In 
Fig.~\ref{fig:pdfcompare}
we show these two models over-plotted on the PDFs drawn
from the forty-five lensing-planes. 

The model of \cite{2006ApJ...645....1D} is
a better fit to the simulation than the lognormal, especially at 
high surface mass density. Note in that
paper the authors used the first year WMAP parameters, whereas the
present simulation is run with the WMAP 3-year parameters, including
a significantly different $\sigma_8$. The fact that the model still
represents the simulation well suggests that it is quite general.
%\clearpage
%\clearpage
\subsection{Tests for the convergence plane}
As described in \S\ref{Sigma_to_Kappa}, the effective convergence
plane was produced by a two step process. First, we computed
the effective convergence plane by weighting the surface mass
density planes from the simulation with appropriate geometrical
factors. Let us call it the map $\bmm M_1$. This map, therefore,
includes contribution from the large scale structure only out to
the redshift of the farthest TPM plane, $z =  4.05$. Next we added
in the contribution from $z>4.05$, by generating a Gaussian random
realization of the effective convergence from a theoretical power
spectrum (the map $\bmm M_2$). Therefore the final convergence
map is simply $(\bmm M_1 +\bmm M_2)$. It is interesting to compare
the power spectrum of the map $\bmm M_1$ with that expected from
theoretical considerations. Since CMB lensing is most sensitive
to large scale modes, we should make sure that these modes were
realized correctly in our simulated convergence plane. Incidentally,
these scales are also linear to mildly nonlinear. Therefore,
we should expect the power spectrum of the convergence map to
be well replicated by the theoretical prediction at least in the
quasi-linear range of  multipoles ($\ell \lesssim 2000$) where
simple non-linear prescriptions suffice.

%\clearpage
\begin{figure}
\center
\includegraphics[scale=0.37]{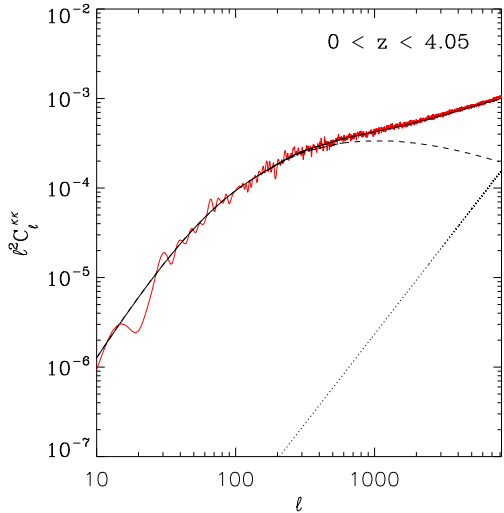}
\caption{\label{fig:Eff_Kappa_to_TPMz} Power spectrum of the
effective convergence map $\bmm M_1$ produced from the simulated lensing
planes alone. The red line shows the power spectrum computed
from the convergence map and the black solid line represents the
theoretical power spectrum with non-linear corrections.  The power spectrum is corrected for the shot noise contribution (see text) which is displayed as the dotted line. The black dashed line corresponds to the linear theory power spectrum. All theory power spectra are mode-coupled with the apodizing window. }
\end{figure}
%\clearpage

In order to compute the theoretical power spectra for the maps $\bmm M_1$ and $(\bmm M_1+\bmm M_2)$, we used the Limber approximation to project 
the matter power
spectrum $P(k, \eta)$ computed from CAMB.
 The Limber approximation simplifies the full curved-sky
calculation, and is valid for $l\gtrsim 10$. Since for lower values
of the multipole we have few realizations of the convergence modes,
the power spectrum computed from the simulated map is noisy in this
regime, rendering it practically useless for comparison with theory. 
Therefore, an accurate computation of the theoretical convergence
power spectrum for these lowest multipoles is unnecessary, and the
Limber calculation suffices. Under the same approximation, the shot noise contribution to the convergence field can be computed as
\beq
\cl^{\mathrm{shot}} = \sum_j \Delta\eta_j \left( \frac32 \Om (1+z_j) \frac{(\eta_s-\eta_j)\eta_j}{\eta_s} \frac{H_0^2}{c^2} \right)^2 \frac{1}{\bar n_j} ,
\eeq
where $\bar n_j = N_j/(\eta_j^2 \Delta\eta_j\tildecaparea)$, $N_j$ being the total number of particles in the $j$-th shell. 

%\clearpage
\begin{figure}
\center
\includegraphics[scale=0.37]{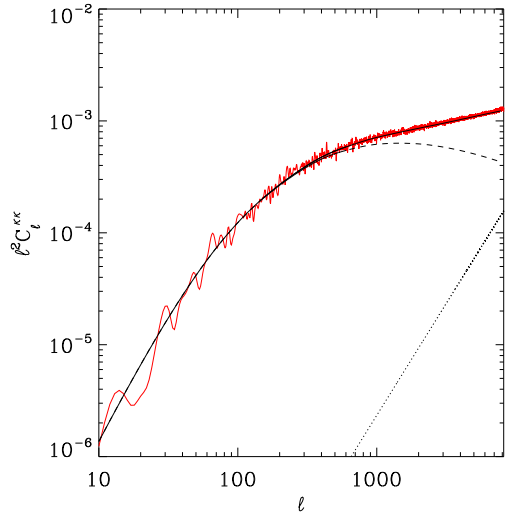}
\caption{\label{fig:Eff_Kappa_to_CMB} Power spectrum of
the effective convergence map $(\bmm M_1+ \bmm M_2)$ after adding in high redshift
contribution. The red line shows the power spectrum computed
from the convergence map and the black solid line represents the
theoretical power spectrum with non-linear corrections. The power spectrum is corrected for the shot noise contribution (see text) which is displayed as the dotted line.  The black dashed line corresponds to the linear theory power spectrum. All theory power spectra are mode-coupled with the apodizing window.}
\end{figure}
%\clearpage

We compute both a linear and a non-linear version of the
convergence power spectrum, where the latter includes non-linear
corrections to the matter power spectrum from a halo model based
fitting formula \citep{2003MNRAS.341.1311S}. We plot the power
spectrum computed from the simulated convergence map $\bmm M_1$,
and the corresponding theoretical power spectra, in 
Fig.~\ref{fig:Eff_Kappa_to_TPMz}. As is evident from the figure, the simulated
power spectrum is in accord with the linear theory power spectrum
for $\ell\lesssim 300$, beyond which the effect non-linearities
creep in. However, it is impressive that the non-linear corrections
to the power spectrum are in good agreement with the simulation
up to relatively high multipoles. The same quantities
are plotted for the convergence map out to the redshift of the CMB
in Fig.~\ref{fig:Eff_Kappa_to_CMB}. 
We find in both cases that beyond multipoles of $\sim 6000$
the simulation contains more non-linear power than predicted by the theory.

\subsection{Tests for the lensed CMB map}\label{Tests_for_lensed_CMB}
Since CMB lensing is essentially a remapping of points, the one-point
statistics should remain unaffected by the lensing.
We check for this by drawing up the one-point PDF's of the unlensed
and lensed maps, and find them to be consistent to within $0.8\%$.
Next, we compare the power spectrum of the simulated lensed map
(cf. Figures~\ref{LensedCMBcls} \& \ref{LensedCMBclsZOOM}) with
the theoretical predictions as computed with the CAMB code. For
a quantitative comparison, we consider a range of  multipoles
$(500 \le \ell\le 3500)$ in the acoustic regime. We do not
consider the lower multipoles as they exhibit negligible lensing
effect. We found that for a fixed input cosmology,
the tail ($\ell\gtrsim 3500$) of the lensed CMB  power spectrum
predicted by CAMB depends somewhat sensitively on input parameters,
specifically the combination $k\eta_\mathrm{max}$, which controls
the maximum value of the wavenumber for which the matter power
spectrum in computed. However, the lensed power spectrum from
CAMB is robust towards changes in the auxiliary input parameters
for the range of multipoles, $\ell < 3500$. Also, the lensed CMB multipoles beyond this range couple to relatively small scale modes of the deflection field where our simulation has more power than expected from non-linear theory. In fact, beyond $\ell \simeq 4000$ the simulated power spectrum is found to deviate systematically from the theoretical spectrum.

As the simulated power spectrum, $\tilde\cl^{\mathrm{sim}}$,
was computed using an apodized window as described in
\S\ref{Angular_Power_Spectra}, the appropriate theoretical curve to
compare this result with is the power spectrum from CAMB after it has
been convolved with the coupling matrix defined by the same weighting
scheme (cf. equation \ref{pseudocl} ).	We denote the latter quantity
by $\tilde {C}_l^{\mathrm{theory}}$. 
In order to facilitate the comparison
we bin the raw spectrum  in $\ell$. In the multipole range considered
$(500 \le \ell\le 4000)$, the quantity ${\cal C}_\ell = \ell^4 \cl$
is flat (see Fig.~\ref{LensedCMBclsZOOM}) and therefore a better
candidate for binning.  We denote the difference between the simulated and
the theoretical version of this quantity by
\beq
\delta{\mathcal C}_l \equiv \tilde{\mathcal C}_l^{\mathrm{sim}} - 
  \tilde {\mathcal C}_l^{\mathrm{theory}}   .
\eeq   
For each of the $N$ bins indexed by $b$, we compute the mean, $\delta{\mathcal C}_b$, and the sample variance, $s^2_b$, of the observations falling inside that bin. In order to account for that fact that cosmic variance errors will be higher in our case due to incomplete sky coverage, we define an effective variance as  $\sigma_b^2 = s_b^2/\fsky^{\mathrm{eff}}$.  

%\clearpage
\begin{figure}
\center
\includegraphics[scale=0.37]{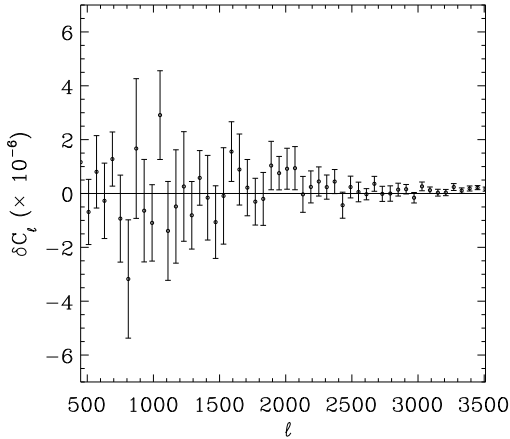}
\caption{\label{fig:Error_Analysis} Difference between the
simulated and the theoretical binned power spectrum for lensed CMB.}
\end{figure}
%\clearpage

 We quantify the goodness of fit between the simulation and the model
by defining a $\chi^2$ statistic as
\beq
\chi^2 = \sum_{b=1}^{N} \frac{({\delta\mathcal C}_b)^{2}}{\sigma_b^2}.
\eeq
 We perform the $\chi^2$ analysis by uniformly binning the power
 spectra in the range $500 \le \ell\le 4000$ into $52$ bins with a
 bin width of $\Delta \ell = 60$.The binned values 
along with the error bars are displayed in Fig.~\ref{fig:Error_Analysis}.
 We find a value $\chi^2 = 52.93$, suggesting an appreciable agreement with the theory. 
\section{Conclusions}\label{Conclusions}
In this paper, we have put forward an algorithm for end-to-end
simulation of the gravitational lensing of the cosmic microwave
background, starting with an $N$-body simulation and fully taking
into account the curvature of the sky. The method is applicable to
maps of any geometry on the surface of the sphere, including the
whole sky.  Our algorithm includes prescriptions for generating
spherical convergence planes from an N-body light cone and
subsequently ray-tracing through the planes to simulate lensing. The
central feature of the algorithm is the use of a highly accurate
interpolation method that enables sampling of both the deflection fields
on intermediate lensing planes and the unlensed CMB map on an
irregular grid. We have provided a detailed description of both a
multiple plane ray tracing and an effective lensing version
of the algorithm. The latter setting has been used to illustrate
the algorithm, by generating an $\sim$1$\arcmin$ resolution lensed CMB
map. We have compared the power spectra of the effective convergence
map and the lensed CMB map with theoretical predictions, and have
obtained good agreement. After this paper was completed, \citet{2007arXiv0711.1540F} described a similar method of producing convergence maps in spherical geometry, and \cite{2007arXiv0711.2655C} also described their techniques for simulating CMB lensing maps. The latter used broadly similar techniques to those described here, although they used a different method to obtain the deflection field. 

Applications of the algorithm can be manifold.  The associated
large scale structure planes can be populated with tracers of mass
and foreground sources, in order to simulate cross-correlation studies and
to investigate the effects of contamination. This lensing portion of the algorithm
 can be applied to generate lensed maps in large scale structure simulations 
that produce spherical shells \citep{2007arXiv0711.1540F}. The multiple plane algorithm can
be particularly useful, after trivial modifications, in simulating
weak lensing of galaxies or the 21-cm background on a large sky.
The lensed CMB maps can be used as inputs to telescope simulators
for projects such as ACT and PLANCK, and will help in the analysis and
interpretation of data. We intend to release all-sky high resolution 
lensed CMB maps made using this algorithm in near future.

\acknowledgements
SD would sincerely like to thank his advisor,
David Spergel for suggesting the key ideas of the project and for his
continuous guidance and encouragement throughout its development. SD
is specially grateful to Chris Hirata for generously providing the
NISHT code and for numerous useful discussions. We 
thank Joanna Dunkley for careful reading of the manuscript
and thoughtful suggestions. We would also like to thank the referee for
his thoughtful suggestions. SD acknowledges the support from 
the NASA Theory Program NNG04GK55G and the NSF grant AST-0707731.
This research was facilitated by allocations of advanced
computing resources from the Pittsburgh Supercomputing Center
and the National Center for Supercomputing Applications.
In addition, computational facilities at
Princeton supported by NSF grant AST-0216105 were used, as well as
high performance computational facilities supported by
Princeton University under the auspices of the
Princeton Institute for Computational Science and Engineering
(PICSciE) and the Office of Information Technology (OIT). Some of the
 results in this paper have been derived using the HEALPix \citep{gorski-2005-622} package.

\end{document}